\documentclass[12pt]{article}
\pdfoutput=1

\usepackage{amsmath}
\usepackage{amssymb}
\usepackage{epsfig}
\usepackage{epstopdf}
\usepackage{latexsym}
\usepackage{color}
\numberwithin{equation}{section}
\usepackage[
      colorlinks=false,
      linkcolor=darkblue,  
      urlcolor=blue,    
      filecolor=blue,     
      citecolor=red,
linktocpage=true,
      pdfstartview=FitV,
      bookmarksopen=true    
      ]{hyperref}

\begin{document}

\begin{titlepage}

\centerline{\Huge \rm } 
\bigskip
\centerline{\Huge \rm On-shell action and the Bekenstein-Hawking entropy}
\bigskip
\centerline{\Huge \rm of supersymmetric black holes in $AdS_6$}
\bigskip
\bigskip
\bigskip
\bigskip
\bigskip
\bigskip
\centerline{\rm Minwoo Suh}
\bigskip
\centerline{\it Department of Physics, Kyungpook National University, Daegu 41566, Korea}
\bigskip
\centerline{\tt minwoosuh1@gmail.com} 
\bigskip
\bigskip
\bigskip
\bigskip
\bigskip
\bigskip
\bigskip
\bigskip
\bigskip
\bigskip
\bigskip
\bigskip

\begin{abstract}
\noindent Recently, entropy of static supersymmetric black holes in $AdS_6$ was microscopically counted by the topologically twisted index of five-dimensional superconformal field theories. However, the AdS/CFT dictionary states that the partition function of the boundary field theory corresponds to the on-shell action of the bulk quantum gravity. In this paper, we aim to explain the microscopic counting of black hole entropy by the topologically twisted index. We calculate the renormalized on-shell action and show that the on-shell action equals the Bekenstein-Hawking entropy of the supersymmetric black holes in $AdS_6$.
\end{abstract}

\vskip 5cm

\flushleft {December, 2018}

\end{titlepage}

\tableofcontents

\section{Introduction and conclusions}

Recently, via the AdS/CFT correspondence, \cite{Maldacena:1997re, Gubser:1998bc, Witten:1998qj}, there has been great success in microscopic counting of the Bekenstein-Hawking entropy of supersymmetric black holes in $AdS_4$, \cite{Cacciatori:2009iz, DallAgata:2010ejj, Hristov:2010ri}, by a topologically twisted index of three-dimensional quantum field theories, \cite{Benini:2015noa, Benini:2015eyy}. Further progress has been made for supersymmetric black holes in $AdS_6$, \cite{Suh:2018tul, Hosseini:2018usu, Suh:2018szn}, by the topologically twisted index of five-dimensional superconformal field theories, \cite{Hosseini:2018uzp, Crichigno:2018adf}.

However, the AdS/CFT dictionary, \cite{Maldacena:1997re, Gubser:1998bc, Witten:1998qj}, states that the partition function of the boundary field theory corresponds to the on-shell action of the bulk quantum gravity,
\begin{equation}
-\log{Z}_{\text{CFT}}\,=\,S_{\text{gravity}}\,.
\end{equation}
In this paper, we show that the on-shell action equals the Bekenstein-Hawking entropy of supersymmetric black holes by holographic renormalization, \cite{Bianchi:2001de, Bianchi:2001kw}. We closely follow the parallel calculation performed for supersymmetric black holes in $AdS_4$ in \cite{Halmagyi:2017hmw}. See \cite{Azzurli:2017kxo, Cabo-Bizet:2017xdr} for the related works in $AdS_4$.

In addition, we compute the on-shell actions of the supersymmetric black strings and branes which are interpolating solutions between the boundary of $AdS_6$ and the horizon of $AdS_3\,\times\,H^3$ and $AdS_4\,\times\,\Sigma_{\mathfrak{g}}$, respectively, \cite{Nunez:2001pt}. The on-shell actions of the black strings and branes reproduce the central charge and the three-sphere free energy of 2d and 3d field theories, respectively, \cite{Bobev:2017uzs}.

We make some comments on our results. First, we do not need the explicit form of the black hole solutions and the boundary asymptotics will be enough. Second, the supersymmetry of the black hole solutions is essential. The calculation uses the supersymmetry equations in the course of holographic renormalization to obtain the on-shell action. Third, we find that the counterterms precisely match the counterterms for general solutions of $F(4)$ gauged supergravity derived in \cite{Alday:2014rxa, Alday:2014bta}. They were already employed for calculations of supersymmetric Renyi entropy in \cite{Alday:2014fsa, Hama:2014iea}. It gives a nontrivial verification of counterterms we employ. Fourth, we calculate the on-shell actions of supersymmetric $AdS_6$ black holes, strings and branes. It reveals a uniform structure of on-shell actions of black objects obtained from topological twist in $F(4)$ gauged supergravity, \cite{Romans:1985tw}.

In section 2, we present $F(4)$ gauged supergravity in the Euclidean spacetime. In section 3, for the supersymmetric $AdS_6$ black hole solution from $F(4)$ gauged supergravity, \cite{Suh:2018tul}, we calculate the renormalized on-shell action and prove that it equals the Bekenstein-Hawking entropy of the black holes. In section 4 and 5, we calculate the on-shell actions of supersymmetric black string and brane solutions, \cite{Nunez:2001pt}, and show that they reproduce the central charge and the three-sphere free energy of 2d and 3d dual field theories, respectively. In appendix A, we review the counterterms for pure $AdS$ gravity theories in diverse dimensions. In appendix B, we review the counterterms for $F(4)$ gauged supergravity in \cite{Alday:2014rxa, Alday:2014bta}.

\section{Euclidean $F(4)$ gauged supergravity}

We present $F(4)$ gauged supergravity, \cite{Romans:1985tw}, in Euclidean spacetime. The Euclidean continuation is performed by a Wick rotation,
\begin{equation} \label{wickt}
t\,\rightarrow\,i\tau\,.
\end{equation}
As the signature of $F(4)$ gauged supergravity in \cite{Romans:1985tw} is $(+-----)$, we obtain the Euclidean theory in the signature of $(------)$.  It accompanies the Wick rotation of the time component of gamma matrices,
\begin{equation} \label{wickgt}
\gamma_t\,\rightarrow\,-i\gamma_\tau\,,
\end{equation}
which renders the anti-commutation relations to be
\begin{equation}
\{\gamma_\mu,\gamma_\nu\}\,=\,2\delta_{\mu\nu}\,.
\end{equation}
According to the Euclidean continuation, we employ the chirality operator to be
\begin{equation}
\gamma_7\,=\,-\gamma_{\hat{t}}\gamma_{\hat{x}_1}\gamma_{\hat{x}_2}\gamma_{\hat{x}_3}\gamma_{\hat{x}_4}\gamma_{\hat{x_5}}\,\rightarrow\,i\gamma_{\hat{\tau}}\gamma_{\hat{x}_1}\gamma_{\hat{x}_2}\gamma_{\hat{x}_3}\gamma_{\hat{x}_4}\gamma_{\hat{x_5}}\,,
\end{equation}
where the hatted indices are of the flat spacetime. Also the time components of the fields are Wick rotated, $e.g.$,
\begin{equation}
A_t\,\rightarrow\,-iA_\tau\,, \qquad B_{t\mu}\,\rightarrow\,-iB_{\tau\mu}\,.
\end{equation}

The bosonic field content consists of the metric, $g_{\mu\nu}$, a real scalar, $\phi$, an $SU(2)$ gauge field, $A^I_\mu$, $I\,=\,1,\,2,\,3$, a $U(1)$ gauge field, $\mathcal{A}_\mu$, and a two-form field, $B_{\mu\nu}$. The fermionic field content is gravitinos, $\psi_{\mu{i}}$, and dilatinos, $\chi_i$, $i\,=\,1,\,2$. The field strengths are defined by
\begin{align}
\mathcal{F}_{\mu\nu}\,=&\,\partial_\mu\mathcal{A}_\nu-\partial_\nu\mathcal{A}_\mu\,, \notag \\
F^I_{\mu\nu}\,=&\,\partial_\mu{A}^I_\nu-\partial_\nu{A}^I_\mu+\tilde{g}\epsilon^{IJK}A^J_\mu{A}^K_\nu\,, \notag \\
G_{\mu\nu\rho}\,=&\,3\partial_{[\mu}B_{\nu\rho]}\,, \notag \\
\mathcal{H}_{\mu\nu}\,=&\,\mathcal{F}_{\mu\nu}+mB_{\mu\nu}\,.
\end{align}
Implicitly implying \eqref{wickt} and \eqref{wickgt}, the apparent expressions of the action, equations of motion, and supersymmetry variations in Euclidean spacetime are identical to the Lorentzian ones up to some subtleties we describe. The bosonic Lagrangian is given by
\begin{align}
e^{-1}\mathcal{L}\,=\,&-\frac{1}{4}R+\frac{1}{2}\partial_\mu\phi\partial^\mu\phi+\frac{1}{8}\left(\tilde{g}^2e^{\sqrt{2}\phi}+4\tilde{g}me^{-\sqrt{2}\phi}-m^2e^{-3\sqrt{2}\phi}\right) \notag \\
&-\frac{1}{4}e^{-\sqrt{2}\phi}\left(\mathcal{H}_{\mu\nu}\mathcal{H}^{\mu\nu}+F^I_{\mu\nu}F^{I\mu\nu}\right)+\frac{1}{12}e^{2\sqrt{2}\phi}G_{\mu\nu\rho}G^{\mu\nu\rho} \notag \\
&-\frac{i}{8}\epsilon^{\mu\nu\rho\sigma\tau\kappa}B_{\mu\nu}\left(\mathcal{F}_{\rho\sigma}\mathcal{F}_{\tau\kappa}+mB_{\rho\sigma}\mathcal{F}_{\tau\kappa}+\frac{1}{3}m^2B_{\rho\sigma}B_{\tau\kappa}+F^I_{\rho\sigma}F^I_{\tau\kappa}\right)\,,
\end{align}
where $\tilde{g}$ is the $SU(2)$ gauge coupling and $m$ is the mass parameter of the two-form field. The expressions are identical to the Lorentzian supergravity, however, note the factor of $i$ in the Chern-Simons terms, \cite{Alday:2014rxa, Alday:2014bta}. It is required for gauge invariance in the path integral with Euclidean measure. It is also implied by supersymmetry. 

The scalar potential,
\begin{equation}
V\,=\,-\frac{1}{8}\left(\tilde{g}^2e^{\sqrt{2}\phi}+4\tilde{g}me^{-\sqrt{2}\phi}-m^2e^{-3\sqrt{2}\phi}\right)\,,
\end{equation}
can be obtained from a superpotential,
\begin{equation} \label{sp}
W\,=\,\frac{1}{2\sqrt{2}}\left(\tilde{g}e^{\frac{\phi}{\sqrt{2}}}+me^{-\frac{3\phi}{\sqrt{2}}}\right)\,,
\end{equation}
from
\begin{equation}
V\,=\,\frac{1}{2}\left(\frac{{\partial}W}{\partial\phi}\right)^2-\frac{5}{4}W^2\,.
\end{equation}

We present the equations of motion,
\begin{align}
&R_{\mu\nu}\,=\,2\partial_\mu\phi\partial_\nu\phi+\frac{1}{8}g_{\mu\nu}\left(\tilde{g}^2e^{\sqrt{2}\phi}+4\tilde{g}me^{-\sqrt{2}\phi}-m^2e^{-3\sqrt{2}\phi}\right)-2e^{-\sqrt{2}\phi}\left(\mathcal{H}_\mu\,^\rho\mathcal{H}_{\nu\rho}-\frac{1}{8}g_{\mu\nu}\mathcal{H}_{\rho\sigma}\mathcal{H}^{\rho\sigma}\right) \notag \\ 
& \,\,\,\,\,\,\,\,\,\,\,\,\,\,\,\,\,\,\,\,\,\,\,\,\,\, -2e^{-\sqrt{2}\phi}\left(F^I_\mu\,^\rho{F}^I_{\nu\rho}-\frac{1}{8}g_{\mu\nu}F^I_{\rho\sigma}F^{I\rho\sigma}\right)+e^{2\sqrt{2}\phi}\left(G_\mu\,^{\rho\sigma}G_{\nu\rho_\sigma}-\frac{1}{6}g_{\mu\nu}G_{\rho\sigma\tau}G^{\rho\sigma\tau}\right)\,, \\
&\frac{1}{\sqrt{-g}}\partial_\mu\left(\sqrt{-g}g^{\mu\nu}\partial_\nu\phi\right)\,=\,\frac{1}{4\sqrt{2}}\left(\tilde{g}^2e^{\sqrt{2}\phi}-4\tilde{g}me^{-\sqrt{2}\phi}+3m^2e^{-3\sqrt{2}\phi}\right) \notag \\ 
& \,\,\,\,\,\,\,\,\,\,\,\,\,\,\,\,\,\,\,\,\,\,\,\,\,\, +\frac{1}{2\sqrt{2}}e^{-\sqrt{2}\phi}\left(\mathcal{H}_{\mu\nu}\mathcal{H}^{\mu\nu}+F^I_{\mu\nu}F^{I\mu\nu}\right)+\frac{1}{3\sqrt{2}}e^{2\sqrt{2}\phi}G_{\mu\nu\rho}G^{\mu\nu\rho}\,, \\
&\mathcal{D}_\nu\left(e^{-\sqrt{2}\phi}\mathcal{H}^{\nu\mu}\right)\,=\,\frac{i}{6}e\epsilon^{\mu\nu\rho\sigma\tau\kappa}\mathcal{H}_{\nu\rho}G_{\sigma\tau\kappa}\,, \\
&\mathcal{D}_\nu\left(e^{-\sqrt{2}\phi}F^{I\nu\mu}\right)\,=\,\frac{i}{6}e\epsilon^{\mu\nu\rho\sigma\tau\kappa}F^I_{\nu\rho}G_{\sigma\tau\kappa}\,, \\
&\mathcal{D}_\rho\left(e^{2\sqrt{2}\phi}G^{\rho\mu\nu}\right)\,=\,-\frac{i}{4}e\epsilon^{\mu\nu\rho\sigma\tau\kappa}\left(\mathcal{H}_{\rho\sigma}\mathcal{H}_{\tau\kappa}+F^I_{\rho\sigma}F^I_{\tau\kappa}\right)-me^{-\sqrt{2}\phi}\mathcal{H}^{\mu\nu}\,.
\end{align}
Note the factors of $i$ in the last three equations which originate from the factor of $i$ in the Chern-Simons terms in the action.

The supersymmetry transformations of the fermionic fields are
\begin{align}
\delta\psi_{\mu{i}}\,=&\,\nabla_\mu\epsilon_i+\tilde{g}A^I_\mu(T^I)_i\,^j\epsilon_j-\frac{1}{8\sqrt{2}}\left(\tilde{g}e^{-\frac{\phi}{\sqrt{2}}}+me^{-\frac{3\phi}{\sqrt{2}}}\right)\gamma_\mu\gamma_7\epsilon_i \notag \\
&-\frac{1}{8\sqrt{2}}e^{-\frac{\phi}{\sqrt{2}}}\left(\mathcal{F}_{\nu\lambda}+mB_{\nu\lambda}\right)\left(\gamma_\mu\,^{\nu\lambda}-6\delta_\mu\,^\nu\gamma^\lambda\right)\epsilon_i \notag \\
&-\frac{1}{4\sqrt{2}}e^{-\frac{\phi}{\sqrt{2}}}F^I_{\nu\lambda}\left(\gamma_\mu\,^{\nu\lambda}-6\delta_\mu\,^\nu\gamma^\lambda\right)\gamma_7(T^I)_i\,^j\epsilon_j \notag \\
&-\frac{1}{24}e^{\sqrt{2}\phi}G_{\nu\lambda\rho}\gamma_7\gamma^{\nu\lambda\rho}\gamma_\mu\epsilon_i\,,
\end{align}
\begin{align}
\delta\chi_i\,=&\,\frac{1}{\sqrt{2}}\gamma^\mu\partial_\mu\phi\epsilon_i+\frac{1}{4\sqrt{2}}\left(\tilde{g}e^{-\frac{\phi}{\sqrt{2}}}-3me^{-\frac{3\phi}{\sqrt{2}}}\right)\gamma_7\epsilon_i \notag \\
&+\frac{1}{4\sqrt{2}}e^{-\frac{\phi}{\sqrt{2}}}\left(\mathcal{F}_{\mu\nu}+mB_{\mu\nu}\right)\gamma^{\mu\nu}\epsilon_i \notag \\
&+\frac{1}{2\sqrt{2}}e^{-\frac{\phi}{\sqrt{2}}}F^I_{\mu\nu}\gamma^{\mu\nu}\gamma_7(T^I)_i\,^j\epsilon_j \notag \\
&-\frac{1}{12}e^{\sqrt{2}\phi}G_{\mu\nu\lambda}\gamma_7\gamma^{\mu\nu\lambda}\epsilon_i\,,
\end{align}
where $T^I$, $I$ = 1, 2, 3, are the $SU(2)$ left-invariant one-forms,
\begin{equation}
T^I\,=\,-\frac{i}{2}\sigma^I\,.
\end{equation}
The expressions are identical to the Lorentzian ones with \eqref{wickt} and \eqref{wickgt} implied implicitly.

\section{On-shell action of $AdS_6$ black holes}

We compute the renormalized on-shell action of the supersymmetric $AdS_6$ black holes from $F(4)$ gauged supergravity, \cite{Suh:2018tul}, and show that it equals the Bekenstein-Hawking entropy of the black holes.

\subsection{Euclidean black holes}

We present the supersymmetric black holes with a horizon which is a product of two Riemann surfaces, \cite{Suh:2018tul}, in Euclidean spacetime. We consider the metric,
\begin{equation}
ds^2\,=\,-e^{2f(r)}\left(dt^2+dr^2\right)-e^{2g_1(r)}\left(d\theta_1^2+\sinh^2\theta_1d\phi_1^2\right)-e^{2g_2(r)}\left(d\theta_2^2+\sinh^2\theta_1d\phi_2^2\right)\,,
\end{equation}
for the $\Sigma_{\mathfrak{g}_1}\times\Sigma_{\mathfrak{g}_2}$ background with $\mathfrak{g}_1>0$ and $\mathfrak{g}_2>0$. The only non-vanishing component of the non-Abelian $SU(2)$ gauge field, $A^I_\mu$, $I$ = 1, 2, 3, is given by
\begin{equation}
A^3\,=\,a_1\cosh\theta_1{d}\phi_1+a_2\cosh\theta_2{d}\phi_2\,,
\end{equation}
where the magnetic charges, $a_1$ and $a_2$, are constant. We turn off the $U(1)$ gauge field, $\mathcal{A}_\mu$.

The final expressions of the supersymmetry equations are identical to the Lorentzian ones. Therefore, we briefly summarize the differences in the derivations. The two-form field is determined by solving the equations of motion,
\begin{equation} \label{twoform}
B_{tr}\,=\,-\frac{2i}{m^2}a_1a_2e^{\sqrt{2}\phi+2f-2g_1-2g_2}\,,
\end{equation}
where there is a factor of $i$ in contrast to the Lorentzian solution. As discussed before, time components of the fields are Wick rotated. In the derivation of supersymmetry equations, this factor of $i$ is canceled by the $i$ in the time component of gamma matrices. We present the supersymmetry equations,
\begin{align} \label{susye}
f'e^{-f}\,=&\,-\frac{1}{4\sqrt{2}}\left(\tilde{g}e^{\frac{\phi}{\sqrt{2}}}+me^{-\frac{3\phi}{\sqrt{2}}}\right)+\frac{k}{2\sqrt{2}\tilde{g}}e^{-\frac{\phi}{\sqrt{2}}}\left(e^{-2g_1}+e^{-2g_2}\right)-\frac{3}{\sqrt{2}\tilde{g}^2m}e^{\frac{\phi}{\sqrt{2}}-2g_1-2g_2}\,, \notag \\ 
g_1'e^{-f}\,=&\,-\frac{1}{4\sqrt{2}}\left(\tilde{g}e^{\frac{\phi}{\sqrt{2}}}+me^{-\frac{3\phi}{\sqrt{2}}}\right)-\frac{k}{2\sqrt{2}\tilde{g}}e^{-\frac{\phi}{\sqrt{2}}}\left(3e^{-2g_1}-e^{-2g_2}\right)+\frac{1}{\sqrt{2}\tilde{g}^2m}e^{\frac{\phi}{\sqrt{2}}-2g_1-2g_2}\,, \notag \\ 
g_2'e^{-f}\,=&\,-\frac{1}{4\sqrt{2}}\left(\tilde{g}e^{\frac{\phi}{\sqrt{2}}}+me^{-\frac{3\phi}{\sqrt{2}}}\right)-\frac{k}{2\sqrt{2}\tilde{g}}e^{-\frac{\phi}{\sqrt{2}}}\left(3e^{-2g_2}-e^{-2g_1}\right)+\frac{1}{\sqrt{2}\tilde{g}^2m}e^{\frac{\phi}{\sqrt{2}}-2g_1-2g_2}\,, \notag \\ 
\frac{1}{\sqrt{2}}\phi'e^{-f}\,=&\,\frac{1}{4\sqrt{2}}\left(\tilde{g}e^{\frac{\phi}{\sqrt{2}}}-3me^{-\frac{3\phi}{\sqrt{2}}}\right)-\frac{k}{2\sqrt{2}\tilde{g}}e^{-\frac{\phi}{\sqrt{2}}}\left(e^{-2g_1}+e^{-2g_2}\right)-\frac{1}{\sqrt{2}\tilde{g}^2m}e^{\frac{\phi}{\sqrt{2}}-2g_1-2g_2}\,,
\end{align}
with
\begin{equation}
a_1\,=\,-\frac{k}{\lambda\tilde{g}}\,, \qquad a_2\,=\,-\frac{k}{\lambda\tilde{g}}\,,
\end{equation}
where $\lambda\,=\,\pm1$ and $k\,=\,-1$ for the $\Sigma_{\mathfrak{g}_1}\times\Sigma_{\mathfrak{g}_2}$ background with $\mathfrak{g}_1>0$ and $\mathfrak{g}_2>0$. When checking that the equations of motion are satisfied by the supersymmetry equations, factors of $-1$ from products of two-form fields are canceled by the change of time component of the metric, $g_{tt}\,\rightarrow\,-g_{tt}$.

\subsection{On-shell action}

\subsubsection{The bulk and boundary terms}

In this section, we calculate the on-shell action. The calculation is quite parallel to the calculation of on-shell action of $AdS_4$ black holes in \cite{Halmagyi:2017hmw}. The Einstein equations for the ansatz in the previous section can be presented by
\begin{align}
0\,=&\,e^{-2f}R_{tt}+V+F+3B\,, \notag \\
0\,=&\,e^{-2f}R_{rr}+4\Phi-V-F-3B\,, \notag \\
0\,=&\,\frac{1}{2}\left(e^{-2g_1}R_{\theta_1\theta_1}+e^{-2g_2}R_{\theta_2\theta_2}\right)-V+F+B\,,
\end{align}
where we define
\begin{align}
V\,=&\,-\frac{1}{8}\left(\tilde{g}^2e^{\sqrt{2}\phi}+4\tilde{g}me^{-\sqrt{2}\phi}-m^2e^{-3\sqrt{2}\phi}\right)\,, \notag \\
F\,=&\,-\frac{1}{2}e^{-\sqrt{2}\phi}\left(a_1^2e^{-4g_1}+a_2^2e^{-4g_2}\right)\,, \notag \\
B\,=&\,-\frac{2}{m^2}a_1^2a_2^2e^{\sqrt{2}\phi-4g_1-4g_2}\,, \notag \\
\Phi\,=&\,-\frac{1}{2}e^{-2f}\phi'\phi'\,,
\end{align}
and $V$ is the scalar potential. With the Einstein equations, the bosonic Lagrangian can be expressed in terms of the metric functions,
\begin{align}
e^{-1}\mathcal{L}\,&=\,-\frac{1}{4}R+\Phi-V+F+B \notag \\
&=\,-\frac{1}{4}\left(R+e^{-2f}\left(R_{tt}+R_{rr}\right)+2\left(e^{-2g_1}R_{\theta_1\theta_1}+e^{-2g_2}R_{\theta_2\theta_2}\right)\right) \notag \\
&=\,-\frac{1}{2}e^{-2f}\left(f''+2f'\left(g_1'+g_2'\right)\right) \notag \\
&=\,-\frac{1}{4}e^{-2f-2g_1-2g_2}\left(e^{-2f+2g_1+2g_2}\left(e^{2f}\right)'\right)'\,.
\end{align}
The bulk action and the Gibbons-Hawking term, \cite{Gibbons:1976ue, York:1972sj}, are, respectively,
\begin{align}
S_{bulk}\,&=\,\frac{1}{4{\pi}G_N}{\int}d^6x\mathcal{L} \notag \\
&=\,\frac{{\beta}vol(\Sigma_{\mathfrak{g}_1})vol(\Sigma_{\mathfrak{g}_2})}{8{\pi}G_N}\int_{r_h}^{r_0}dr\frac{1}{2}\left(e^{-2f+2g_1+2g_2}\left(e^{2f}\right)'\right)'\,, \\
S_{GH}\,&=\,-\frac{1}{8\pi{G}_N}\int_\partial{d}^5x\sqrt{h}\mathcal{K} \notag \\
&=-\left.\frac{{\beta}vol(\Sigma_{\mathfrak{g}_1})vol(\Sigma_{\mathfrak{g}_2})}{8{\pi}G_N}e^{2g_1+2g_2}\left(f'+2g_1'+2g_2'\right)\right|_{r=r_0}\,,
\end{align}
where $h_{ij}$ is the induced metric. The extrinsic curvature, normal vector, and trace of the extrinsic curvature are, respectively,
\begin{equation}
\mathcal{K}_{ij}\,=\,\frac{1}{2}n^k\partial_k{g}_{ij}\,, \qquad n^j\,=\,\frac{1}{\sqrt{g_{rr}}}\left(\frac{\partial}{\partial{r}}\right)^j\,, \qquad \mathcal{K}\,=\,h^{ij}\mathcal{K}_{ij}\,.
\end{equation}
We denote the six-dimensional Newton's gravitational constant by $G_N$. The radius of the horizon and the location of the boundary are denoted by $r_h$ and $r_0$, respectively. We obtain the inverse temperature by
\begin{equation}
\beta\,=\,\frac{1}{T}\,=\,\frac{4\pi}{e^{2f}\left(-e^{-2f}\right)'}\,.
\end{equation}
Then the sum of the bulk action and the Gibbons-Hawking term is 
\begin{equation}
S_{bulk}+S_{GH}\,=\,\left.\frac{A}{4G_N}-\frac{{\beta}vol(\Sigma_{\mathfrak{g}_1})vol(\Sigma_{\mathfrak{g}_2})}{8{\pi}G_N}\left(e^{2g_1+2g_2}\right)'\right|_{r=r_0}\,,
\end{equation}
where the area of the horizon is
\begin{equation}
A\,=\,\left.e^{2g_1+2g_2}vol(\Sigma_{\mathfrak{g}_1})vol(\Sigma_{\mathfrak{g}_2})\right|_{r=r_h}\,.
\end{equation}
Employing the supersymmetry equations in \eqref{susye}, we can rewrite
\begin{align} \label{expr}
\left(e^{2g_1+2g_2}\right)'\,&=\,2e^{2g_1+2g_2}\left(g_1'+g_2'\right) \notag \\
&=4e^{f+2g_1+2g_2}\left[-\frac{1}{2}W+\frac{\lambda}{2\sqrt{2}}e^{-\frac{\phi}{\sqrt{2}}}\left(a_1e^{-2g_1}+a_2e^{-2g_2}\right)+\frac{1}{\sqrt{2}m}a_1a_2e^{\frac{\phi}{\sqrt{2}}-2g_1-2g_2}\right]\,,
\end{align}
where $W$ is the superpotential in \eqref{sp}. Using the expression in \eqref{expr}, the sum of the bulk action and the Gibbons-Hawking term is
\begin{align} \label{abcd}
S_{bulk}+&S_{GH}\,=\,\left.\frac{A}{4G_N}+\frac{{\beta}vol(\Sigma_{\mathfrak{g}_1})vol(\Sigma_{\mathfrak{g}_2})}{4{\pi}G_N}e^{f+2g_1+2g_2}W\right|_{r=r_0} \notag \\
-&\left.\frac{{\beta}vol(\Sigma_{\mathfrak{g}_1})vol(\Sigma_{\mathfrak{g}_2})}{2{\pi}G_N}e^{f+2g_1+2g_2}\left[\frac{\lambda}{2\sqrt{2}}e^{-\frac{\phi}{\sqrt{2}}}\left(a_1e^{-2g_1}+a_2e^{-2g_2}\right)+\frac{1}{\sqrt{2}m}a_1a_2e^{\frac{\phi}{\sqrt{2}}-2g_1-2g_2}\right]\right|_{r=r_0}\,.
\end{align}
The full black hole solution is an interpolating geometry between the asymptotic $AdS_6$ boundary at $r_0\,=\,0$ and the $AdS_2\,\times\,H^2\,\times\,H^2$ horizon at $r_h\,=\,\infty$. At the boundary, $r_0\,=\,0$, the metric and the scalar field asymptote to $AdS_6$ spacetime,
\begin{equation}
e^f\,=\,e^{g_1}\,=\,e^{g_2}\,\cong\,\frac{R_{AdS_6}}{r}\,, \qquad e^{\frac{\phi}{\sqrt{2}}}\,\cong\,1\,.
\end{equation}
Hence, we observe that the superpotential term, the term with $a_1$ and $a_2$, the term with $a_1a_2$ in \eqref{abcd} are quintic, cubic, and linear divergences, respectively, at the boundary, $r=r_0$. Therefore, in the next subsection, we will introduce counterterms to eliminate the divergent terms in the on-shell action.

\subsubsection{The counterterms}

In order to eliminate the divergent terms in the on-shell action, we introduce the counterterms. Accordingly, we introduce the superpotential counterterm and counterterms from the curvature of the boundary, respectively,{\footnote{We are grateful to Ioannis Papadimitriou for explaining the $\mathcal{R}^2$ counterterms in $AdS_6$.}}
\begin{align}
S_{SUSY}\,&=\,-\frac{1}{8{\pi}G_N}\int_{\partial}d^5x\sqrt{h}\Big[2W\Big] \notag \\
&=-\left.\frac{{\beta}vol(\Sigma_{\mathfrak{g}_1})vol(\Sigma_{\mathfrak{g}_2})}{4{\pi}G_N}e^{f+2g_1+2g_2}W\right|_{r=r_0}\,, \\ \label{rr}
S_{\mathcal{R}}\,&=\,-\frac{1}{8{\pi}G_N}\int_{\partial}d^5x\sqrt{h}\left[\frac{1}{\sqrt{2}\tilde{g}}e^{-\frac{\phi}{\sqrt{2}}}\mathcal{R}\right] \notag \\
&=-\left.\frac{{\beta}vol(\Sigma_{\mathfrak{g}_1})vol(\Sigma_{\mathfrak{g}_2})}{2{\pi}G_N}e^{f+2g_1+2g_2}\left[\frac{k}{2\sqrt{2}\tilde{g}}e^{-\frac{\phi}{\sqrt{2}}}\left(e^{-2g_1}+e^{-2g_2}\right)\right]\right|_{r=r_0}\,, \\ \label{RR}
S_{\mathcal{R}^2}\,&=\,-\frac{1}{8{\pi}G_N}\int_{\partial}d^5x\sqrt{h}\left[\frac{2}{\sqrt{2}\tilde{g}^2m}e^{\frac{\phi}{\sqrt{2}}}\left(\mathcal{R}_{ij}\mathcal{R}^{ij}-\frac{5}{16}\mathcal{R}^2\right)\right] \notag \\
&=\left.\frac{{\beta}vol(\Sigma_{\mathfrak{g}_1})vol(\Sigma_{\mathfrak{g}_2})}{4{\pi}G_N}e^{f+2g_1+2g_2}\left[\frac{1}{\sqrt{2}\tilde{g}^2m}e^{\frac{\phi}{\sqrt{2}}-4g}\right]\right|_{r=r_0}\,,
\end{align}
where the Ricci scalar of the induced metric is
\begin{equation}
\mathcal{R}\,=\,2k\left(e^{-2g_1}+e^{-2g_2}\right)\,.
\end{equation}
In \eqref{RR} we set
\begin{equation}
g\,\equiv\,g_1\,=\,g_2
\end{equation}
which is indeed the case for supersymmetric solutions. Combining the counterterms to the bulk action and the Gibbons-Hawking term, we have
\begin{align} \label{eqeq}
S_{\text{on-shell}}\,&=\,S_{bulk}+S_{GH}+S_{SUSY}+S_{\mathcal{R}}+S_{\mathcal{R}^2} \notag \\
&=\,\left.\frac{A}{4G_N}-\frac{{\beta}vol(\Sigma_{\mathfrak{g}_1})vol(\Sigma_{\mathfrak{g}_2})}{4{\pi}G_N}e^{f+4g}\left[\frac{1}{\sqrt{2}m}a_1a_2e^{\frac{\phi}{\sqrt{2}}-4g}\right]\right|_{r=r_0}\,.
\end{align}
Therefore, the counterterms we introduced have eliminated the quintic and cubic divergences and the half of the linear divergence in the on-shell action. We still have the second term in \eqref{eqeq} which is linearly divergent at the boundary.

If we set the scalar field to vanish, $\phi\,=\,0$, and use $\tilde{g}\,=\,3m$, we obtain
\begin{align}
S_{SUSY}+S_{\mathcal{R}}+S_{\mathcal{R}^2}\,&=\,\frac{1}{8{\pi}G_N}\int_{\partial}d^5x\sqrt{h}\left[\frac{4}{3\sqrt{2}}\tilde{g}+\frac{1}{\sqrt{2}\tilde{g}}\mathcal{R}+\frac{6}{\sqrt{2}g^3}\left(\mathcal{R}_{ij}\mathcal{R}^{ij}-\frac{5}{16}\mathcal{R}^2\right)\right]\,, \notag \\
&=\frac{1}{8{\pi}G_N}\int_{\partial}d^5x\sqrt{h}\left[\frac{4}{l}+\frac{l}{6}\mathcal{R}+\frac{l^3}{18}\left(\mathcal{R}_{ij}\mathcal{R}^{ij}-\frac{5}{16}\mathcal{R}^2\right)\right]\,,
\end{align}
where, in the second line, we reparametrized $\tilde{g}\rightarrow3\sqrt{2}/l$. This precisely coincides with the counterterms introduced for pure gravity in $AdS_6$ in \eqref{ejmct}. This provides a non-trivial check of our counterterms against the counterterms of pure gravity in $AdS$.

Returning to the on-shell action, in order to eliminate the second term in \eqref{eqeq}, we introduce additional counterterms,
\begin{align} \label{ctf}
S_{F^2}\,=&\,\frac{1}{8{\pi}G_N}\int_{\partial}d^5x\sqrt{h}\frac{1}{\sqrt{2}m}F^I_{ij}F^{Iij}\,, \\ \label{ctph}
S_{\phi}\,=&\,\frac{1}{8{\pi}G_N}\int_{\partial}d^5x\sqrt{h}\left(-\frac{1}{2}\right)\frac{4\tilde{g}}{3\sqrt{2}}\left(1-e^\frac{\phi}{\sqrt{2}}\right)^2\,,
\end{align}
which we obtain from the counterterms for general solutions of $F(4)$ gauged supergravity derived in \cite{Alday:2014rxa, Alday:2014bta}. See appendix B for details. For the counterterm, $S_\phi$, we have an additional factor of $-1/2$ compared to the one in \cite{Alday:2014rxa, Alday:2014bta}. The factor of this term is freely determined to cancel the divergences. See, $e.g.$, explanation around (4.32) of \cite{Gutperle:2018axv}. Asymptotically solving the supersymmetry equations, \eqref{susye}, around the boundary, $r\,\rightarrow\,0$, we obtain the solutions,
\begin{align}
f\,=&\,-\log\frac{r}{l}-\frac{1}{18}r^2+\cdots\,, \notag \\
g_1\,=&\,-\log\frac{r}{l}+\frac{1}{9}r^2+\cdots\,, \notag \\
g_2\,=&\,-\log\frac{r}{l}+\frac{1}{9}r^2+\cdots\,, \notag \\
\phi\,=&\,-\frac{1}{3\sqrt{2}}r^2+\cdots\,,
\end{align}
where $l\,=\,3\sqrt{2}/\tilde{g}$ is the radius of $AdS_6$. By plugging the asymptotic solutions into the counterterms, \eqref{ctf} and \eqref{ctph}, we observe that they indeed eliminate the divergences in the second term of \eqref{eqeq}. 

We finally obtain that the renormalized on-shell action,
\begin{align} \label{fos}
S_{\text{on-shell}}\,&=\,S_{bulk}+S_{GH}+S_{SUSY}+S_{\mathcal{R}}+S_{\mathcal{R}^2}+S_{F^2}+S_{\phi}=\,\frac{A}{4G_N}.
\end{align}
which equals the Bekenstein-Hawking entropy of the supersymmetric black holes in $AdS_6$.{\footnote{As we are in the Euclidean spacetime of $(------)$ signature, the on-shell action gives the Bekenstein-Hawking entropy, not the minus of it.}} The counterterms precisely match the counterterms for general solutions of $F(4)$ gauged supergravity derived in \cite{Alday:2014rxa, Alday:2014bta}, which we present in appendix B.{\footnote{As our two-form field, \eqref{twoform}, has a component on the radial direction, $r$, its pull-back on the induced metric trivially vanishes. Therefore, all terms involving the two-form field in \eqref{f4ct} identically vanishes.}}

\section{On-shell action of $AdS_6$ black strings}

We compute the renormalized on-shell action of the supersymmetric $AdS_6$ black strings from $F(4)$ gauged supergravity, \cite{Nunez:2001pt}, and show that it reproduces the central charge of 2d dual conformal field theories.

\subsection{Euclidean black strings}

We present the supersymmetric black strings, \cite{Nunez:2001pt},  in Euclidean spacetime. It is obtained from D4-branes wrapped on a compact hyperbolic space, $H^3$, which is an interpolating solution between the boundary of $AdS_6$ and the horizon of $AdS_3\,\times\,H^3$. We consider the metric,
\begin{equation} \label{bbmet3}
ds^2\,=\,-e^{2f(r)}\left(dt^2+du^2+dr^2\right)-e^{2g(r)}\left(d\varphi^2+\sinh^2\varphi{d}\theta^2+\sinh^2\varphi\sin^2\theta{d}\psi^2\right)\,.
\end{equation}
The non-vanishing components of the non-Abelian $SU(2)$ gauge field, $A^I_\mu$, $I$ = 1, 2, 3, are given by
\begin{equation}
A^1\,=\,a\cosh\varphi{d}\theta\,, \qquad A^2\,=\,b\cos\theta{d}\psi\,, \qquad A^3\,=\,c\sin\theta\cosh\varphi{d}\psi\,,
\end{equation}
where the magnetic charges, $a$, $b$ and $c$, are constant. We turn off the $U(1)$ gauge field, $\mathcal{A}_\mu$, and the two-form field, $B_{\mu\nu}$.

We present the supersymmetry equations,
\begin{align} \label{susye3}
f'e^{-f}\,=&\,-\frac{1}{4\sqrt{2}}\left(\tilde{g}e^{\frac{\phi}{\sqrt{2}}}+me^{-\frac{3\phi}{\sqrt{2}}}\right)+\frac{3k}{2\sqrt{2}\tilde{g}}e^{-\frac{\phi}{\sqrt{2}}-2g}\,, \notag \\ 
g'e^{-f}\,=&\,-\frac{1}{4\sqrt{2}}\left(\tilde{g}e^{\frac{\phi}{\sqrt{2}}}+me^{-\frac{3\phi}{\sqrt{2}}}\right)-\frac{5k}{2\sqrt{2}\tilde{g}}e^{-\frac{\phi}{\sqrt{2}}-2g}\,, \notag \\ 
\frac{1}{\sqrt{2}}\phi'e^{-f}\,=&\,\frac{1}{4\sqrt{2}}\left(\tilde{g}e^{\frac{\phi}{\sqrt{2}}}-3me^{-\frac{3\phi}{\sqrt{2}}}\right)-\frac{3k}{2\sqrt{2}\tilde{g}}e^{-\frac{\phi}{\sqrt{2}}-2g}\,,
\end{align}
with the twist condition,
\begin{equation}
a\,=\,-\frac{k}{\tilde{g}}\,, \qquad b\,=\,-\frac{k}{\tilde{g}}\,, \qquad c\,=\,\frac{k}{\tilde{g}}\,,
\end{equation}
where $k\,=\,-1$ for the $H^3$ background.

\subsection{On-shell action}

\subsubsection{The bulk and boundary terms}

In this section, we calculate the on-shell action. The calculation is quite parallel to the calculation of the on-shell action of $AdS_4$ black holes in \cite{Halmagyi:2017hmw}. The Einstein equations for the ansatz in the previous section can be presented by
\begin{align}
0\,=&\,e^{-2f}R_{tt}-V-F\,, \notag \\
0\,=&\,e^{-2f}R_{rr}+4\Phi-V-F\,, \notag \\
0\,=&\,e^{-2g}R_{\varphi\varphi}-V+\frac{5}{3}F\,,
\end{align}
where we define
\begin{align}
V\,=&\,-\frac{1}{8}\left(\tilde{g}^2e^{\sqrt{2}\phi}+4\tilde{g}me^{-\sqrt{2}\phi}-m^2e^{-3\sqrt{2}\phi}\right)\,, \notag \\
F\,=&\,-\frac{3}{2g^2}e^{-\sqrt{2}\phi-4g}\,, \notag \\
\Phi\,=&\,-\frac{1}{2}e^{-2f}\phi'\phi'\,,
\end{align}
and $V$ is the scalar potential. With the Einstein equations, the bosonic Lagrangian can be expressed in terms of the metric functions,
\begin{align}
e^{-1}\mathcal{L}\,&=\,-\frac{1}{4}R+\Phi-V+F \notag \\
&=\,-\frac{1}{4}R-\frac{1}{4}\left(e^{-2f}R_{rr}+3e^{-2g}R_{\varphi\varphi}\right) \notag \\
&=\,-\frac{1}{2}e^{-2f}\left(f''+f'f'+3f'g'\right) \notag \\
&=\,-\frac{1}{2}e^{-3f-3g}\left(e^{3g}\left(e^f\right)'\right)'\,.
\end{align}
The bulk action and the Gibbons-Hawking term, \cite{Gibbons:1976ue, York:1972sj}, are, respectively,
\begin{align}
S_{bulk}\,&=\,\frac{1}{4{\pi}G_N}{\int}d^6x\mathcal{L} \notag \\
&=\,l_u\frac{{\beta}vol(H^3)}{8{\pi}G_N}\int_{r_h}^{r_0}dr\left(e^{3g}\left(e^f\right)'\right)'\,, \\
S_{GH}\,&=\,-\frac{1}{8\pi{G}_N}\int_\partial{d}^5x\sqrt{h}\mathcal{K} \notag \\
&=-\left.l_u\frac{{\beta}vol(H^3)}{8{\pi}G_N}e^{f+3g}\left(2f'+3g'\right)\right|_{r=r_0}\,,
\end{align}
where $h_{ij}$ is the induced metric and $l_u$ is from the integration over the $u$ coordinate of the metric, \eqref{bbmet3}. The extrinsic curvature, normal vector, and trace of the extrinsic curvature are, respectively,
\begin{equation}
\mathcal{K}_{ij}\,=\,\frac{1}{2}n^k\partial_k{g}_{ij}\,, \qquad n^j\,=\,\frac{1}{\sqrt{g_{rr}}}\left(\frac{\partial}{\partial{r}}\right)^j\,, \qquad \mathcal{K}\,=\,h^{ij}\mathcal{K}_{ij}\,.
\end{equation}
We denote the six-dimensional Newton's gravitational constant by $G_N$. The radius of the horizon and the location of the boundary are denoted by $r_h$ and $r_0$, respectively. We obtain the inverse temperature by
\begin{equation}
\beta\,=\,\frac{1}{T}\,=\,\frac{4\pi}{e^{2f}\left(-e^{-2f}\right)'}\,.
\end{equation}
Then the sum of the bulk action and the Gibbons-Hawking term is 
\begin{equation}
S_{bulk}+S_{GH}\,=\,\left.l_u\frac{e^{f+3g}vol(H^3)}{4G_N}\right|_{r=r_h}-\left.l_u\frac{{\beta}vol(H^3)}{8{\pi}G_N}\left(e^{f+3g}\right)'\right|_{r=r_0}\,.
\end{equation}
Employing the supersymmetry equations in \eqref{susye3}, we can rewrite
\begin{align} \label{expr3}
\left(e^{f+3g}\right)'\,&=\,e^{f+3g}\left(f'+3g'\right) \notag \\
&=4e^{2f+3g}\left[-\frac{1}{2}W+\frac{3k}{2\sqrt{2}\tilde{g}}e^{-\frac{\phi}{\sqrt{2}}-2g}\right]\,.
\end{align}
where $W$ is the superpotential in \eqref{sp}. Using the expression in \eqref{expr3}, the sum of the bulk action and the Gibbons-Hawking term is
\begin{align} \label{abcd3}
S_{bulk}+S_{GH}\,=&\,\left.l_u\frac{e^{f+3g}vol(H^3)}{4G_N}\right|_{r=r_h} \notag \\
+&\left(l_u\frac{{\beta}vol(H^3)}{4{\pi}G_N}e^{2f+3g}W-\left.l_u\frac{{\beta}vol(H^3)}{2{\pi}G_N}e^{2f+3g}\frac{3k}{2\sqrt{2}\tilde{g}}e^{-\frac{\phi}{\sqrt{2}}-2g}\right)\right|_{r=r_0}\,.
\end{align}
The full solution is an interpolating geometry between the asymptotic $AdS_6$ boundary at $r_0\,=\,0$ and the $AdS_3\,\times\,H^3$ horizon at $r_h\,=\,\infty$. At the boundary, $r_0\,=\,0$, the metric and the scalar field asymptote to $AdS_6$ spacetime,
\begin{equation}
e^f\,=\,e^{g}\,\cong\,\frac{R_{AdS_6}}{r}\,, \qquad e^{\frac{\phi}{\sqrt{2}}}\,\cong\,1\,.
\end{equation}
Hence, we observe that the superpotential term and the term with $1/\tilde{g}$ in \eqref{abcd3} are quintic and cubic divergences, respectively, at the boundary, $r=r_0$. Therefore, in the next subsection, we will introduce counterterms to eliminate the divergent terms in the on-shell action.

\subsubsection{The counterterms}

In order to eliminate the divergent terms in the on-shell action, we introduce the counterterms. Accordingly, we introduce the superpotential counterterm and counterterms from the curvature of the boundary, respectively,
\begin{align}
S_{SUSY}\,&=\,-\frac{1}{8{\pi}G_N}\int_{\partial}d^5x\sqrt{h}\Big[2W\Big] \notag \\
&=-\left.l_u\frac{{\beta}vol(H^3)}{4{\pi}G_N}e^{2f+3g}W\right|_{r=r_0}\,, \\ \label{rr}
S_{\mathcal{R}}\,&=\,-\frac{1}{8{\pi}G_N}\int_{\partial}d^5x\sqrt{h}\left[\frac{1}{\sqrt{2}\tilde{g}}e^{-\frac{\phi}{\sqrt{2}}}\mathcal{R}\right] \notag \\
&=-\left.l_u\frac{{\beta}vol(H^3)}{2{\pi}G_N}e^{2f+3g}\left[\frac{3k}{2\sqrt{2}\tilde{g}}e^{-\frac{\phi}{\sqrt{2}}-2g}\right]\right|_{r=r_0}\,, \\
S_{\mathcal{R}^2}\,&=\,-\frac{1}{8{\pi}G_N}\int_{\partial}d^5x\sqrt{h}\left[\frac{2}{\sqrt{2}\tilde{g}^2m}e^{\frac{\phi}{\sqrt{2}}}\left(\mathcal{R}_{ij}\mathcal{R}^{ij}-\frac{5}{16}\mathcal{R}^2\right)\right] \notag \\
&=\left.l_u\frac{{\beta}vol(H^3)}{4{\pi}G_N}e^{2f+3g}\left[\frac{3}{2\sqrt{2}\tilde{g}^2m}e^{\frac{\phi}{\sqrt{2}}-4g}\right]\right|_{r=r_0}\,,
\end{align}
where the Ricci scalar of the induced metric is
\begin{equation}
\mathcal{R}\,=\,6ke^{-2g}\,.
\end{equation}
Combining the counterterms to the bulk action and the Gibbons-Hawking term, we have
\begin{align} \label{eqeq3}
S_{\text{on-shell}}\,&=\,S_{bulk}+S_{GH}+S_{SUSY}+S_{\mathcal{R}}+S_{\mathcal{R}^2} \notag \\
&=\,\left.l_u\frac{e^{f+3g}vol(H^3)}{4G_N}\right|_{r=r_h}+S_{\mathcal{R}^2}\,.
\end{align}
Therefore, the counterterms we introduced have eliminated the quintic and cubic divergences. However, they also introduced the $S_{\mathcal{R}^2}$ term which diverges linearly.

In order to eliminate the $S_{\mathcal{R}^2}$ term in \eqref{eqeq3}, we introduce additional counterterms,
\begin{align} \label{ctf3}
S_{F^2}\,=&\,\frac{1}{8{\pi}G_N}\int_{\partial}d^5x\sqrt{h}\frac{1}{\sqrt{2}m}F^I_{ij}F^{Iij}\,, \\ \label{ctph3}
S_{\phi}\,=&\,\frac{1}{8{\pi}G_N}\int_{\partial}d^5x\sqrt{h}\left(-\frac{1}{2}\right)\frac{4\tilde{g}}{3\sqrt{2}}\left(1-e^\frac{\phi}{\sqrt{2}}\right)^2\,,
\end{align}
which we obtain from the counterterms for general solutions of $F(4)$ gauged supergravity derived in \cite{Alday:2014rxa, Alday:2014bta}. See appendix B for details. For the counterterm, $S_\phi$, we have an additional factor of $-1/2$ compared to the one in \cite{Alday:2014rxa, Alday:2014bta}. The factor of this term is freely determined to cancel the divergences. See, $e.g.$, explanation around (4.32) of \cite{Gutperle:2018axv}. Asymptotically solving the supersymmetry equations, \eqref{susye3}, around the boundary, $r\,\rightarrow\,0$, we obtain the solutions,
\begin{align}
f\,=&\,-\log\frac{r}{l}-\frac{1}{12}r^2+\cdots\,, \notag \\
g\,=&\,-\log\frac{r}{l}+\frac{1}{4}r^2+\cdots\,, \notag \\
\phi\,=&\,-\frac{1}{2\sqrt{2}}r^2+\cdots\,,
\end{align}
where $l\,=\,3\sqrt{2}/\tilde{g}$ is the radius of $AdS_6$. By plugging the asymptotic solutions into the counterterms, \eqref{ctf3} and \eqref{ctph3}, we observe that they indeed eliminate the divergences in the second term of \eqref{eqeq3}. 

We finally obtain the renormalized on-shell action,
\begin{align} \label{fos3}
S_{\text{on-shell}}\,&=\,S_{bulk}+S_{GH}+S_{SUSY}+S_{\mathcal{R}}+S_{\mathcal{R}^2}+S_{F^2}+S_{\phi}=\,\left.l_u\frac{e^{f+3g}vol(H^3)}{4G_N}\right|_{r=r_h}.
\end{align}
The on-shell action per unit volume of $u$ space, \eqref{bbmet}, reproduces the central charge of 2d dual conformal field theories,
\begin{equation}
\frac{S_{\text{on-shell}}}{V}\,=\,\left.\frac{e^{f+3g}vol(H^3)}{4G_N}\right|_{r=r_h}\,=\,\frac{c_{2d}}{6},
\end{equation}
where $V\,=\,l_u$. See, $e.g.$, (4.34), (B.3) and (B.5) in \cite{Bobev:2017uzs}. The counterterms precisely match the counterterms for general solutions of $F(4)$ gauged supergravity derived in \cite{Alday:2014rxa, Alday:2014bta}, which we present in appendix B.

\section{On-shell action of $AdS_6$ black branes}

We compute the renormalized on-shell action of the supersymmetric $AdS_6$ black branes from $F(4)$ gauged supergravity, \cite{Nunez:2001pt}, and show that it reproduces the three-sphere free energy of 3d dual field theories.

\subsection{Euclidean black branes}

We present the supersymmetric black branes with a horizon of a Riemann surface, \cite{Nunez:2001pt},  in Euclidean spacetime. It is obtained from D4-branes wrapped on a Riemann surface which is an interpolating solution between the boundary of $AdS_6$ and the horizon of $AdS_4\,\times\,\Sigma_{\mathfrak{g}}$. It was studied from ten-dimensional supergravity perspective in \cite{Bah:2018lyv} and from field theory in \cite{Crichigno:2018adf}. We consider the metric,
\begin{equation} \label{bbmet}
ds^2\,=\,-e^{2f(r)}\left(dt^2+dx^2+dy^2+dr^2\right)-e^{2g(r)}\left(d\theta^2+\sinh^2\theta{d}\varphi^2\right)\,,
\end{equation}
for the $\Sigma_{\mathfrak{g}}$ background with $\mathfrak{g}\,>0\,$. The only non-vanishing component of the non-Abelian $SU(2)$ gauge field, $A^I_\mu$, $I$ = 1, 2, 3, is given by
\begin{equation}
A^3\,=\,a\cosh\theta{d}\varphi\,,
\end{equation}
where the magnetic charge, $a$, is constant. We turn off the $U(1)$ gauge field, $\mathcal{A}_\mu$, and the two-form field, $B_{\mu\nu}$.

We present the supersymmetry equations,
\begin{align} \label{susye2}
f'e^{-f}\,=&\,-\frac{1}{4\sqrt{2}}\left(\tilde{g}e^{\frac{\phi}{\sqrt{2}}}+me^{-\frac{3\phi}{\sqrt{2}}}\right)+\frac{k}{2\sqrt{2}\tilde{g}}e^{-\frac{\phi}{\sqrt{2}}-2g}\,, \notag \\ 
g'e^{-f}\,=&\,-\frac{1}{4\sqrt{2}}\left(\tilde{g}e^{\frac{\phi}{\sqrt{2}}}+me^{-\frac{3\phi}{\sqrt{2}}}\right)-\frac{k}{2\sqrt{2}\tilde{g}}e^{-\frac{\phi}{\sqrt{2}}-2g}\,, \notag \\ 
\frac{1}{\sqrt{2}}\phi'e^{-f}\,=&\,\frac{1}{4\sqrt{2}}\left(\tilde{g}e^{\frac{\phi}{\sqrt{2}}}-3me^{-\frac{3\phi}{\sqrt{2}}}\right)-\frac{k}{2\sqrt{2}\tilde{g}}e^{-\frac{\phi}{\sqrt{2}}-2g}\,,
\end{align}
with
\begin{equation}
a\,=\,-\frac{k}{\lambda\tilde{g}}\,,
\end{equation}
where $\lambda\,=\,\pm1$ and $k\,=\,-1$ for the $\Sigma_{\mathfrak{g}}$ background with $\mathfrak{g}\,>0\,$.

\subsection{On-shell action}

\subsubsection{The bulk and boundary terms}

In this section, we calculate the on-shell action. The calculation is quite parallel to the calculation of the on-shell action of $AdS_4$ black holes in \cite{Halmagyi:2017hmw}. The Einstein equations for the ansatz in the previous section can be presented by
\begin{align}
0\,=&\,e^{-2f}R_{tt}-V-F\,, \notag \\
0\,=&\,e^{-2f}R_{rr}+4\Phi-V-F\,, \notag \\
0\,=&\,e^{-2g}R_{\theta\theta}-V+3F\,,
\end{align}
where we define
\begin{align}
V\,=&\,-\frac{1}{8}\left(\tilde{g}^2e^{\sqrt{2}\phi}+4\tilde{g}me^{-\sqrt{2}\phi}-m^2e^{-3\sqrt{2}\phi}\right)\,, \notag \\
F\,=&\,-\frac{1}{2}a^2e^{-\sqrt{2}\phi-4g}\,, \notag \\
\Phi\,=&\,-\frac{1}{2}e^{-2f}\phi'\phi'\,,
\end{align}
and $V$ is the scalar potential. With the Einstein equations, the bosonic Lagrangian can be expressed in terms of the metric functions,
\begin{align}
e^{-1}\mathcal{L}\,&=\,-\frac{1}{4}R+\Phi-V+F \notag \\
&=\,-\frac{1}{4}R+\frac{1}{4}e^{-2f}\left(R_{tt}-R_{rr}\right)-\frac{1}{2}\left(e^{-2f}R_{tt}+e^{-2g}R_{\theta\theta}\right) \notag \\
&=\,-\frac{1}{2}e^{-2f}\left(f''+2f'f'+2f'g'\right) \notag \\
&=\,-\frac{1}{4}e^{-4f-2g}\left(e^{2g}\left(e^{2f}\right)'\right)'\,.
\end{align}
The bulk action and the Gibbons-Hawking term, \cite{Gibbons:1976ue, York:1972sj}, are, respectively,
\begin{align}
S_{bulk}\,&=\,\frac{1}{4{\pi}G_N}{\int}d^6x\mathcal{L} \notag \\
&=\,l_xl_y\frac{{\beta}vol(\Sigma_{\mathfrak{g}})}{8{\pi}G_N}\int_{r_h}^{r_0}dr\frac{1}{2}\left(e^{2g}\left(e^{2f}\right)'\right)'\,, \\
S_{GH}\,&=\,-\frac{1}{8\pi{G}_N}\int_\partial{d}^5x\sqrt{h}\mathcal{K} \notag \\
&=-\left.l_xl_y\frac{{\beta}vol(\Sigma_{\mathfrak{g})}}{8{\pi}G_N}e^{2f+2g}\left(3f'+2g'\right)\right|_{r=r_0}\,,
\end{align}
where $h_{ij}$ is the induced metric and $l_x$ and $l_y$ are from the integration over $x$ and $y$ coordinates of the metric, \eqref{bbmet}. The extrinsic curvature, normal vector, and trace of the extrinsic curvature are, respectively,
\begin{equation}
\mathcal{K}_{ij}\,=\,\frac{1}{2}n^k\partial_k{g}_{ij}\,, \qquad n^j\,=\,\frac{1}{\sqrt{g_{rr}}}\left(\frac{\partial}{\partial{r}}\right)^j\,, \qquad \mathcal{K}\,=\,h^{ij}\mathcal{K}_{ij}\,.
\end{equation}
We denote the six-dimensional Newton's gravitational constant by $G_N$. The radius of the horizon and the location of the boundary are denoted by $r_h$ and $r_0$, respectively. We obtain the inverse temperature by
\begin{equation}
\beta\,=\,\frac{1}{T}\,=\,\frac{4\pi}{e^{2f}\left(-e^{-2f}\right)'}\,.
\end{equation}
Then the sum of the bulk action and the Gibbons-Hawking term is 
\begin{equation}
S_{bulk}+S_{GH}\,=\,\left.l_xl_y\frac{e^{2f+2g}vol(\Sigma_{\mathfrak{g}})}{4G_N}\right|_{r=r_h}-\left.l_xl_y\frac{{\beta}vol(\Sigma_{\mathfrak{g}})}{8{\pi}G_N}\left(e^{2f+2g}\right)'\right|_{r=r_0}\,.
\end{equation}
Employing the supersymmetry equations in \eqref{susye2}, we can rewrite
\begin{align} \label{expr2}
\left(e^{2f+2g}\right)'\,&=\,2e^{2f+2g}\left(f'+g'\right) \notag \\
&=4e^{3f+2g}\left[-\frac{1}{2}W+\frac{\lambda}{2\sqrt{2}\tilde{g}}e^{-\frac{\phi}{\sqrt{2}}-2g}\right]\,.
\end{align}
where $W$ is the superpotential in \eqref{sp}. Using the expression in \eqref{expr2}, the sum of the bulk action and the Gibbons-Hawking term is
\begin{align} \label{abcd2}
S_{bulk}+S_{GH}\,=&\,\left.l_xl_y\frac{e^{2f+2g}vol(\Sigma_{\mathfrak{g}})}{4G_N}\right|_{r=r_h} \notag \\ 
+&\left(l_xl_y\frac{{\beta}vol(\Sigma_{\mathfrak{g}})}{4{\pi}G_N}e^{3f+2g}W-\left.l_xl_y\frac{{\beta}vol(\Sigma_{\mathfrak{g}})}{2{\pi}G_N}e^{3f+2g}\frac{\lambda}{2\sqrt{2}\tilde{g}}e^{-\frac{\phi}{\sqrt{2}}-2g}\right)\right|_{r=r_0}\,.
\end{align}
The full solution is an interpolating geometry between the asymptotic $AdS_6$ boundary at $r_0\,=\,0$ and the $AdS_4\,\times\,\Sigma_{\mathfrak{g}}$ horizon at $r_h\,=\,\infty$. At the boundary, $r_0\,=\,0$, the metric and the scalar field asymptote to $AdS_6$ spacetime,
\begin{equation}
e^f\,=\,e^{g}\,\cong\,\frac{R_{AdS_6}}{r}\,, \qquad e^{\frac{\phi}{\sqrt{2}}}\,\cong\,1\,.
\end{equation}
Hence, we observe that the superpotential term and the term with $1/\tilde{g}$ in \eqref{abcd2} are quintic and cubic divergences, respectively, at the boundary, $r=r_0$. Therefore, in the next subsection, we will introduce counterterms to eliminate the divergent terms in the on-shell action.

\subsubsection{The counterterms}

In order to eliminate the divergent terms in the on-shell action, we introduce the counterterms. Accordingly, we introduce the superpotential counterterm and counterterms from the curvature of the boundary, respectively,
\begin{align}
S_{SUSY}\,&=\,-\frac{1}{8{\pi}G_N}\int_{\partial}d^5x\sqrt{h}\Big[2W\Big] \notag \\
&=-\left.l_xl_y\frac{{\beta}vol(\Sigma_{\mathfrak{g}})}{4{\pi}G_N}e^{3f+2g}W\right|_{r=r_0}\,, \\ \label{rr}
S_{\mathcal{R}}\,&=\,-\frac{1}{8{\pi}G_N}\int_{\partial}d^5x\sqrt{h}\left[\frac{1}{\sqrt{2}\tilde{g}}e^{-\frac{\phi}{\sqrt{2}}}\mathcal{R}\right] \notag \\
&=-\left.l_xl_y\frac{{\beta}vol(\Sigma_{\mathfrak{g}})}{2{\pi}G_N}e^{3f+2g}\left[\frac{k}{2\sqrt{2}\tilde{g}}e^{-\frac{\phi}{\sqrt{2}}-2g}\right]\right|_{r=r_0}\,, \\
S_{\mathcal{R}^2}\,&=\,-\frac{1}{8{\pi}G_N}\int_{\partial}d^5x\sqrt{h}\left[\frac{2}{\sqrt{2}\tilde{g}^2m}e^{\frac{\phi}{\sqrt{2}}}\left(\mathcal{R}_{ij}\mathcal{R}^{ij}-\frac{5}{16}\mathcal{R}^2\right)\right] \notag \\
&=\left.l_xl_y\frac{{\beta}vol(\Sigma_{\mathfrak{g}_1})}{4{\pi}G_N}e^{3f+2g}\left[\frac{3}{2\sqrt{2}\tilde{g}^2m}e^{\frac{\phi}{\sqrt{2}}-4g}\right]\right|_{r=r_0}\,,
\end{align}
where the Ricci scalar of the induced metric is
\begin{equation}
\mathcal{R}\,=\,2ke^{-2g}\,.
\end{equation}
Combining the counterterms to the bulk action and the Gibbons-Hawking term, we have
\begin{align} \label{eqeq2}
S_{\text{on-shell}}\,&=\,S_{bulk}+S_{GH}+S_{SUSY}+S_{\mathcal{R}}+S_{\mathcal{R}^2} \notag \\
&=\,\left.l_xl_y\frac{e^{2f+2g}vol(\Sigma_{\mathfrak{g}})}{4G_N}\right|_{r=r_h}+S_{\mathcal{R}^2}\,.
\end{align}
Therefore, the counterterms we introduced have eliminated the quintic and cubic divergences. However, they also introduced the $S_{\mathcal{R}^2}$ term which diverges linearly.

In order to eliminate the $S_{\mathcal{R}^2}$ term in \eqref{eqeq2}, we introduce additional counterterms,
\begin{align} \label{ctf2}
S_{F^2}\,=&\,\frac{1}{8{\pi}G_N}\int_{\partial}d^5x\sqrt{h}\frac{1}{\sqrt{2}m}F^I_{ij}F^{Iij}\,, \\ \label{ctph2}
S_{\phi}\,=&\,\frac{1}{8{\pi}G_N}\int_{\partial}d^5x\sqrt{h}\left(-\frac{1}{2}\right)\frac{4\tilde{g}}{3\sqrt{2}}\left(1-e^\frac{\phi}{\sqrt{2}}\right)^2\,,
\end{align}
which we obtain from the counterterms for general solutions of $F(4)$ gauged supergravity derived in \cite{Alday:2014rxa, Alday:2014bta}. See appendix B for details. For the counterterm, $S_\phi$, we have an additional factor of $-1/2$ compared to the one in \cite{Alday:2014rxa, Alday:2014bta}. The factor of this term is freely determined to cancel the divergences. See, $e.g.$, explanation around (4.32) of \cite{Gutperle:2018axv}. Asymptotically solving the supersymmetry equations, \eqref{susye2}, around the boundary, $r\,\rightarrow\,0$, we obtain the solutions,
\begin{align}
f\,=&\,-\log\frac{r}{l}-\frac{1}{36}r^2+\cdots\,, \notag \\
g\,=&\,-\log\frac{r}{l}+\frac{5}{36}r^2+\cdots\,, \notag \\
\phi\,=&\,-\frac{1}{6\sqrt{2}}r^2+\cdots\,,
\end{align}
where $l\,=\,3\sqrt{2}/\tilde{g}$ is the radius of $AdS_6$. By plugging the asymptotic solutions into the counterterms, \eqref{ctf2} and \eqref{ctph2}, we observe that they indeed eliminate the divergences in the second term of \eqref{eqeq2}. 

We finally obtain the renormalized on-shell action,
\begin{align} \label{fos2}
S_{\text{on-shell}}\,&=\,S_{bulk}+S_{GH}+S_{SUSY}+S_{\mathcal{R}}+S_{\mathcal{R}^2}+S_{F^2}+S_{\phi}=\,\left.l_xl_y\frac{e^{2f+2g}vol(\Sigma_{\mathfrak{g}})}{4G_N}\right|_{r=r_h}.
\end{align}
The on-shell action per unit volume of $x$ and $y$ space, \eqref{bbmet}, reproduces the three-sphere free energy of 3d dual field theories,
\begin{equation}
\frac{S_{\text{on-shell}}}{V}\,=\,\left.\frac{e^{2f+2g}vol(\Sigma_{\mathfrak{g}})}{4G_N}\right|_{r=r_h}\,=\,\frac{F_{S^3}}{2\pi},
\end{equation}
where $V\,=\,l_xl_y$. See, $e.g.$, (4.31), (B.3) and (B.8) in \cite{Bobev:2017uzs}. The counterterms precisely match the counterterms for general solutions of $F(4)$ gauged supergravity derived in \cite{Alday:2014rxa, Alday:2014bta}, which we present in appendix B.

\bigskip
\bigskip
\leftline{\bf Acknowledgements}
\noindent We would like to thank Hyojoong Kim and Ioannis Papadimitriou for invaluable conversations for this work. We are grateful to Nikolay Bobev for reading a draft of the manuscript. We are also grateful to an anonymous referee for a suggestion to improve the manuscript. This research was supported by the National Research Foundation of Korea under the grant NRF-2017R1D1A1B03034576 and NRF-2019R1I1A1A01060811.

\appendix
\section{Counterterms for pure gravity in $AdS_{n+1}$}
\renewcommand{\theequation}{A.\arabic{equation}}
\setcounter{equation}{0} 

For pure gravity theories in $AdS_{n+1}$ spacetimes with $n<7$, the counterterms to eliminate the divergent terms arising from the on-shell action at the boundary were systematically obtained long ago in \cite{Emparan:1999pm}. The counterterms for generic pure gravity theories in $AdS_{n+1}$ spacetimes with $n<7$, are, \cite{Emparan:1999pm},
\begin{equation}\label{ejmct}
S_{ct}\,=\,\frac{1}{8{\pi}G_N}\int_{\partial}d^nx\sqrt{h}\left[\frac{n-1}{l}+\frac{l}{2(n-2)}\mathcal{R}+\frac{l^3}{2(n-4)(n-2)^2}\left(\mathcal{R}_{ij}\mathcal{R}^{ij}-\frac{n}{4(n-1)}\mathcal{R}^2\right)+\cdots\right]\,,
\end{equation}
where $h$, $\mathcal{R}$, $\mathcal{R}_{ij}$ are the determinant of the metric, the Ricci scalar, and the Ricci tensor of the boundary metric, respectively, and $l$ is the radius of $AdS_{n+1}$. According to \cite{Emparan:1999pm}, the first term first appeared in \cite{Hyun:1998vg}, the second term in \cite{Balasubramanian:1999re}, and the third term in \cite{Emparan:1999pm}. In odd dimensions, some denominators vanish, and they produce the logarithmic corrections, $e.g.$, \cite{Bobev:2013cja, Bobev:2016nua}. In $AdS_4$, as the third term is not divergent and vanishes at the boundary, there are counterterms only up to $\mathcal{R}$, $e.g.$, \cite{Freedman:2013ryh, Freedman:2016yue, Halmagyi:2017hmw, Bobev:2018wbt}. In $AdS_6$ we encounter the $\mathcal{R}^2$ counterterms, as they appeared in \cite{Alday:2014rxa, Alday:2014bta, Gutperle:2018axv}.

\section{Counterterms for $F(4)$ gauged supergravity}
\renewcommand{\theequation}{B.\arabic{equation}}
\setcounter{equation}{0} 

We present the counterterms for general solutions of $F(4)$ gauged supergravity derived in \cite{Alday:2014rxa, Alday:2014bta},
\begin{align} \label{f4ct}
S_{ct}\,=&\,\frac{1}{8\pi{G}_N}\int_\partial\left[\left(\frac{4\sqrt{2}}{3}+\frac{1}{2\sqrt{2}}\mathcal{R}[h]-\frac{1}{6\sqrt{2}}||\tilde{B}||_h^2\right.\right. \notag \\
+&\frac{3}{4\sqrt{2}}\mathcal{R}[h]_{mn}\mathcal{R}[h]^{mn}-\frac{15}{64\sqrt{2}}\mathcal{R}[h]^2 \notag \\
-&\frac{3}{4\sqrt{2}}||\tilde{F}^I||_h^2+\frac{1}{12\sqrt{2}}Tr_h\tilde{B}^4-\frac{13}{192\sqrt{2}}||\tilde{B}||^4_h-\frac{1}{\sqrt{2}}||d\tilde{B}||^2_h \notag \\
+&\frac{5}{8\sqrt{2}}||d*_h\tilde{B}+\frac{i\sqrt{2}}{3}\tilde{B}\wedge{\tilde{B}}||^2_h-\frac{1}{4\sqrt{2}}\langle{\tilde{B}},d\delta_h\tilde{B}+\frac{i\sqrt{2}}{3}d*_h\tilde{B}\wedge{\tilde{B}}\rangle_h \notag \\
+&\left.\frac{4\sqrt{2}}{3}\left(1-X\right)^2-\frac{1}{\sqrt{2}}\langle\text{Ric}[h]\circ{\tilde{B}},\tilde{B}\rangle_h+\frac{9}{32\sqrt{2}}\mathcal{R}[h]||\tilde{B}||^2_h\right)d^5x\sqrt{h}\notag \\
-&\left.\frac{1}{4\sqrt{2}}\tilde{B}\wedge\left(d*_hd\tilde{B}+\frac{i\sqrt{2}}{3}\tilde{B}\wedge\delta_h\tilde{B}-\frac{2}{9}\tilde{B}\wedge*_h(\tilde{B}\wedge{\tilde{B}})\right)\right]\,,
\end{align}
where it is defined to be
\begin{equation} \label{defsq}
||F^I||^2_h\,=\,\frac{1}{2!}F^I_{mn}F^{Imn}\,.
\end{equation}
In the first line of \eqref{f4ct}, the first term is quintic and the second and third terms are cubic divergences. The rest of the counterterms are linear divergences.

Beside the definition of \eqref{defsq}, the conventions of \cite{Cvetic:1999un} are employed in \cite{Alday:2014rxa, Alday:2014bta}. They are related to our conventions of \cite{Romans:1985tw} by
\begin{align}
g\,=&\,2\tilde{g}\,, \qquad X\,=\,e^{-\frac{\tilde{\phi}}{2\sqrt{2}}}\,=\,e^{\frac{\phi}{\sqrt{2}}}\,, \notag \\
A^I_\mu\,=&\,\frac{1}{2}\tilde{A}^I_\mu\,, \qquad A_\mu\,=\,\frac{1}{2}\tilde{A}_\mu\,, \qquad B_{\mu\nu}\,=\,\frac{1}{2}\tilde{B}_{\mu\nu}\,,
\end{align}
where the tilded ones are from \cite{Cvetic:1999un}. The signature is chosen to be $(++++++)$ compared to our $(------)$. The gauge coupling and the radius of $AdS_6$ are fixed to be $g\,=\,2$ and $l\,=\,3\sqrt{2}/2$, respectively. 

\vspace{2cm}



\end{document}